\documentclass[conference]{IEEEtran}
\IEEEoverridecommandlockouts
\usepackage{cite}
\usepackage{amsmath,amssymb,amsfonts}
\usepackage{algorithmic}
\usepackage{graphicx}
\usepackage{textcomp}
\usepackage{xcolor}
\usepackage[hyphens]{url}
\usepackage[hidelinks]{hyperref}
\usepackage{booktabs}
\usepackage{float}
\def\BibTeX{{\rm B\kern-.05em{\sc i\kern-.025em b}\kern-.08em
    T\kern-.1667em\lower.7ex\hbox{E}\kern-.125emX}}
\begin{document}

\title{Statistical Assertions for Debugging Quantum Circuits and States in CUDA-Q\\
}

\author{
\IEEEauthorblockN{Jocelyn Li}
\IEEEauthorblockA{Department of Computer Science \\
Princeton University\\
Princeton, NJ, USA \\
jl1543@princeton.edu}
\and

\IEEEauthorblockN{Ella Rubinshtein}
\IEEEauthorblockA{Department of Electrical and 
\\ Computer Engineering \\
Princeton University\\
Princeton, NJ, USA \\
ellarubinshtein@princeton.edu}
\and

\IEEEauthorblockN{Margaret Martonosi}
\IEEEauthorblockA{Department of Computer Science \\
Princeton University\\
Princeton, NJ, USA \\
mrm@princeton.edu}
}

\maketitle

\begin{abstract}
As quantum computing continues to mature, more developers are designing, coding, and simulating quantum circuits. A challenge exists, however, in debugging quantum circuits, particularly as they scale in size and complexity. Given the lack of effective debugging workflows, developers are forced to manually inspect their circuits and analyze various quantum states, which is error-prone and time-consuming.

In this research, we present a statistical assertion-based debugging workflow for CUDA-Q. CUDA-Q has gained popularity due to its ability to leverage GPUs to accelerate quantum circuit simulations; this allows circuits to scale to larger depths and widths, where they can be particularly hard to debug by hand. Inspired by and building from prior Qiskit-based debuggers, our work allows CUDA-Q users to verify quantum program correctness with greater ease. Through the insertion of statistical assertions within a quantum circuit, our tool provides valuable insights into the state of qubits at any point within a circuit, tracks their evolution, and helps detect deviations from expected behavior. Furthermore, we improve the reliability and accuracy of the product state assertion by using a combination of Fisher's exact test and the Monte Carlo Method instead of a chi-square test, and examine the impact of CUDA-Q's distinct kernel-based programming model on the design of our debugging tool. This work offers a practical solution to one of CUDA-Q's usability gaps, paving the way for more reliable and efficient quantum software development.
\end{abstract}


\section{Introduction}
Quantum computing has advanced significantly in recent years, enabling a growing number of developers to design, simulate, and execute quantum circuits. The number of qubits that both physical quantum processors and classical simulators can handle continues to increase. This progress has spurred a broader adoption of quantum-inspired methodologies across disciplines, which is supported by various quantum programming frameworks, including Qiskit and CUDA-Q. These frameworks provide users with tools to develop and execute quantum algorithms on both simulators and quantum processing units (QPUs).

CUDA-Q \cite{nvidia_cudaq_home}, a relatively new quantum development platform created by NVIDIA, has quickly gained traction due to its ability to leverage GPUs to accelerate quantum circuit simulations. This capability allows users to simulate larger and more complex quantum circuits than was previously feasible. However, this increased circuit depth and width leads to increased difficulty in verification and debugging. Quantum programs are already inherently challenging to debug, especially during execution. Unlike classical programs, where programmers can inspect variables at runtime through print statements, quantum states cannot be directly observed. This is because these states can only be observed through measurement, but measurement irreversibly collapses the quantum state from a possibly probabilistic state into a classical value. This makes it difficult to trace the evolution of qubit states throughout a circuit and verify the correctness of intermediate quantum computations.

Additionally, as quantum circuits scale, the state space grows exponentially, further complicating debugging because it becomes very hard to observe and reason about entanglement relationships in the full state space of interesting quantum programs. Finally, the probabilistic nature of quantum measurements adds another layer of complexity to interpreting the measurement output. Therefore, there is a pressing need for effective debugging workflows tailored for quantum programs. 

To address this need, we present a CUDA-Q debugging workflow inspired by and expanding on existing Qiskit-based frameworks. Our approach introduces a practical tool for validating the correctness of CUDA-Q quantum programs. In particular, we investigate how a combination of Fisher's exact test and the Monte Carlo Method can be used to improve the reliability of the product state assertion, and examine the implications of CUDA-Q's distinct kernel-based programming model on our tool. This paper contributes a systematic methodology for debugging CUDA-Q programs, offering insights into the unique challenges of the platform and equipping developers with improved capabilities to test and verify quantum code.

\section{Prior Work}
Previous research \cite{yipengpaper} has introduced statistical assertions as a method for verifying the state of a quantum circuit at various points during execution. Specifically, three types of assertions were  proposed: classical state, uniform state, and product state assertions. These leverage chi-square tests and contingency tables to statistically compare the observed measurement distributions with the expected distributions corresponding to each type of quantum state. 

This statistical assertion methodology was implemented in Qiskit \cite{yipeng_github}, providing developers with tools for debugging Qiskit quantum programs. Our work here builds upon this foundation by analyzing and refining the statistical tests used in the original assertions to improve their accuracy and compatibility with quantum states. Furthermore, we extend this workflow to CUDA-Q, and in doing so, analyze the intricacies and implications of CUDA-Q's kernel-based programming model on our debugging tool. 

\section{Background on Quantum States and CUDA-Q}
\subsection{Quantum States}
Unlike classical computers, which use classical bits that take on the values of either 0 or 1, quantum computers use quantum bits (qubits), which can exist in a probabilistic superposition. Upon measurement, a qubit collapses to a classical value of either 0 or 1, according to the probability distribution determined by its prior quantum state. This collapse upon measurement fundamentally limits a developer’s ability to observe the internal state of a quantum program as it runs on a QPU; this makes traditional debugging techniques, such as printing variable values, inapplicable.

A second fundamental property of quantum systems is entanglement. This is where qubits become correlated in such a way that the state of one qubit is dependent on the state of another. When qubits are entangled, their joint state must be described as a superposition over a larger set of elementary states. This state space grows exponentially with the number of qubits, reflecting both the computational power and the complexity of quantum systems.

This exponential state space growth presents significant challenges for simulation and debugging. In addition to the inability to inspect quantum states without collapsing their state, the sheer size of the state space makes it infeasible for classical systems to fully simulate or visualize the entire quantum state space for even moderately-sized programs. These limitations highlight the pressing need for rigorous and systematic debugging workflows tailored specifically for quantum programs.

\subsection{CUDA-Q}
NVIDIA's CUDA-Q \cite{nvidia_cudaq_home} has emerged as a compelling quantum programming framework, particularly for its ability to smoothly leverage GPU acceleration and thereby significantly enhance the performance of quantum circuit simulation. This capability allows developers to simulate wider and deeper quantum circuits more efficiently, which is a key advantage for algorithm prototyping and testing. However, as we scale to larger circuits, better debugging support is needed.

At the core of CUDA-Q is the quantum kernel, which is the fundamental unit of quantum execution \cite{nvidia_cudaq_kernel_intro}. Quantum kernels are functions that are designed to run on either simulated or real QPUs. These kernels can be executed in conjunction with classical functions, enabling developers to construct flexible, hybrid programs that seamlessly alternate between quantum and classical computation. 

An important distinction within CUDA-Q is that while all quantum circuits are implemented as kernels, not all kernels necessarily represent complete quantum circuits \cite{nvidia_cudaq_kernel_intro}. In particular, quantum circuits can be made up of multiple kernels, with one main kernel encapsulating them all, enabling modular program design. Moreover, since kernels are functions, they can incorporate classical control flow constructs such as conditionals and loops, providing additional flexibility for expressing complex quantum logic. 

\section{Statistical Assertions for Quantum Programs}
Prior work \cite{yipengpaper} has introduced three primary statistical assertions for debugging quantum programs. These assertions analyze the measurement distribution of a quantum state and determine whether it aligns with a specific expected distribution for a quantum circuit. Specifically, these three assertions verify classical, uniform, and product states. Our work not only implements these three assertions in CUDA-Q, but also improves the accuracy of the product state assertion. 

\begin{figure}[t]
  \centering
  \includegraphics[width=\linewidth]{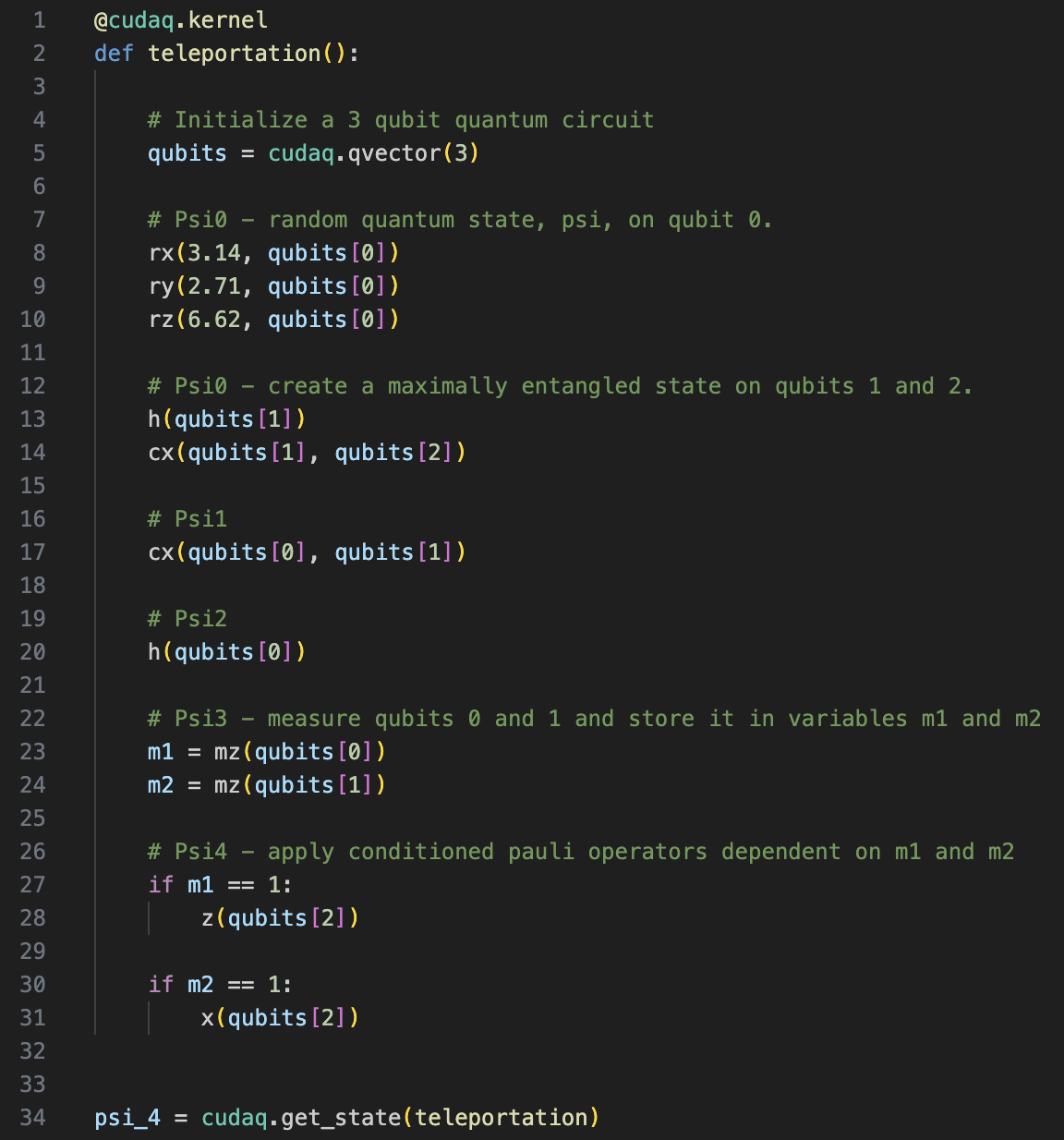}
  \caption{Example implementation of quantum teleportation using a static CUDA-Q kernel.}
  \label{fig:Teleport}
\end{figure}

To obtain a measurement distribution, we call the \texttt{cudaq.sample} function on the current quantum kernel. The number of shots passed into \texttt{cudaq.sample}, which represents the number of times the quantum circuit is executed and measured, is either specified by the user or defaults to 1000, following the CUDA-Q documentation \cite{cudaq_sample}. \texttt{cudaq.sample} executes the given quantum kernel and performs repeated measurements, which over all executions, forms the distribution of measured states for the system.

Once a measurement distribution is obtained, we apply a statistical test to determine whether the observed distribution aligns with what's expected for the given assertion. For example, Figure \ref{fig:Teleport} shows an implementation of the Quantum Teleportation circuit \cite{teleportation}. After line 20, the user expects all three qubits to be maximally entangled. To validate this, one can insert a statistical assertion at line 21 to check that the observed measurement distribution of the current quantum state matches with what's expected of an entangled quantum state. Specifically, we employ a statistical test to check whether these qubits, or groups of qubits, are independent of each other or entangled.

Because these tests are probabilistic, they cannot guarantee correctness; instead, they provide statistical confidence that a qubit or group of qubits is in the desired state. To increase confidence, users can rerun assertions with a higher shot count. Increasing the number of shots leads to an observed distribution that more closely approximates the true probability distribution of the quantum state, according to the Glivenko–Cantelli Theorem \cite{glivenko_cantelli}. As a result, the statistical test will provide a more accurate evaluation of the quantum state.

\subsection{Classical State Assertion}
The first statistical assertion determines whether the current quantum state is a {\em classical} state. A qubit (or register of qubits) is considered to be in a classical state if it deterministically yields a specific classical bitstring upon measurement.

We use a chi-square test to evaluate this assertion. The chi-square test returns a p-value, which represents the probability of obtaining a distribution as extreme as the observed one, assuming the null hypothesis (in this case, the expected distribution) is true \cite{chi_square_test}. A small p-value (typically less than or equal to $0.05$) suggests that the observed distribution is unlikely to have occurred under the null hypothesis, indicating a statistically significant difference between the observed and expected distributions. 

For the classical state assertion, the null hypothesis assumes a unimodal distribution with a single sharp peak at the expected classical bitstring and negligible probability elsewhere. If the chi-square test yields a small p-value, we reject the null hypothesis, concluding that the state is unlikely to be classical. Conversely, a large p-value suggests that the observed measurement distribution is consistent with the expected classical distribution, implying that the state is likely classical.

A common use case of classical state assertions is to check whether qubits have been successfully reset to a classical state. For example, this might be done by calling the inverse of the quantum circuit, either in preparation for future use or to cleanup the workspace. Inserting a classical assertion here checks whether this reset is successful. If this assertion fails, but the quantum circuit code seems correct based on other statistical assertions run on it, there is likely a bug in the inverse circuit code.

\subsection{Uniform State Assertion}
The second statistical assertion checks whether the current quantum state is a {\em uniform} state, meaning it is in an equal superposition over all possible computational basis states. For an $n$-qubit system, this corresponds to a uniform distribution over all $2^n$ basis states. 

Similar to the classical state assertion, we use a chi-square test to evaluate whether the observed measurement distribution matches with the expected uniform distribution. The null hypothesis assumes that each of the $2^n$ measurement outcomes occurs with equal probability of $\frac{1}{2^n}$.

If the quantum state is not in a uniform superposition, the observed measurement distribution will deviate significantly from the expected uniform distribution, and the chi-square test will return a small p-value. Conversely, if the state is indeed in a uniform superposition, the measurement outcomes will be approximately evenly distributed, given a sufficient number of shots, resulting in a high p-value and acceptance of the null hypothesis.

Many quantum programs \cite{teleportation} set the input qubits in a uniform superposition state, in preparation for complex quantum state manipulation. Therefore, placing a uniform state assertion after this setup helps to ensure that there are no bugs in it. This is particularly valuable for preventing setup bugs from propagating through a program.

\subsection{Entanglement and Product State Assertion}
The third statistical assertion checks whether the current quantum state is a {\em product} state. Two qubits are in a product state if their joint state can be expressed as the tensor product of their individual states, implying no entanglement \cite{product_state}. Conversely, if qubits are not in a product state, they are likely entangled, meaning their states are strongly correlated.

\subsubsection{Product State Assertion Workflow}
Our product state assertion workflow begins by constructing a contingency table from the observed measurement distribution, based on the two qubits (or groups of qubits) under evaluation. The table dimensions are $2^{num 
\_qubits\_0}$ by $2^{num 
\_qubits\_1}$, where $num\_qubits\_0$ and $num\_qubits\_1$ are the number of qubits in the two groups of qubits being examined, respectively. For example, for the product state assertion on Line 23 in Figure \ref{fig:screenshotBV}, the contingency table constructed would be 32 by 2, because we are checking for a product state between the 5-qubit secret string and the 1-qubit auxillary qubit q. 

Then, unlike prior work, we pass the contingency table into either Fisher's exact test or the Monte Carlo Method. Specifically, we call \texttt{scipy.stats.fisher\_exact(cont\_table)} \cite{scipyfisher}. This method calls Fisher's exact test if the contingency table is of size $2 \times 2$, or else it calls the Monte Carlo Method. Fisher's exact test evaluates the exact probability of obtaining the observed contingency table (or one more extreme), under the null hypothesis that the two qubits are statistically independent \cite{fisher_exact_test}. In other words, it tests whether they are in a product state. A p-value is then computed from this exact probability. The Monte Carlo Method takes in an observed contingency table, and generates many other contingency tables with the same row and column sums under the assumption of statistical independence between the two groups of qubits being examined \cite{montecarlo}. Then, it evaluates how often the generated tables are at least as extreme as the observed table, under the assumption of independence. Based on this, a p-value is calculated.

Finally, we interpret the returned p-value. A small p-value suggests the qubits are statistically dependent, indicating entanglement. Conversely, a large p-value implies that the observed distribution is consistent with the null hypothesis of statistical independence, and thus the qubits are likely in a product state. Therefore, our revised workflow is: constructing a contingency table from an observed measurement distribution, passing it into \texttt{scipy.stats.fisher\_exact(cont\_table)}, then interpreting the returned p-value to determine whether the qubits are likely entangled or not. Using a combination of Fisher's exact test and the Monte Carlo method allows the test to handle contingency tables of any size, and offers greater reliability, especially under the data conditions encountered in quantum programs.

\subsubsection{Why a Combination of Fisher's Exact Test and the Monte Carlo Method?}

We choose to use a combination of Fisher's exact test and the Monte Carlo method instead of a chi-square test for this assertion because they are better suited than the chi-square test for small and sparse measurement distributions, which are not uncommon in quantum programs. Unlike the chi-square test, they remain valid and accurate even when cell counts in the expected contingency table are zero or small \cite{fisher_chi}; such values frequently occur for sparse measurement distributions. 

When the contingency table is of size $2\times2$, Fisher's exact test is called. The formula used to calculate the exact probability of obtaining the observed contingency table under the null hypothesis of statistical independence between the 2 qubits is \cite{fisher_exact_test}:

\[
P(T) = \frac{(A + B)! \, (C + D)! \, (A + C)! \, (B + D)!}{N! \, A! \, B! \, C! \, D!}
\]

\begin{figure}[hbt]
  \centering
  \includegraphics[width=0.75\linewidth]{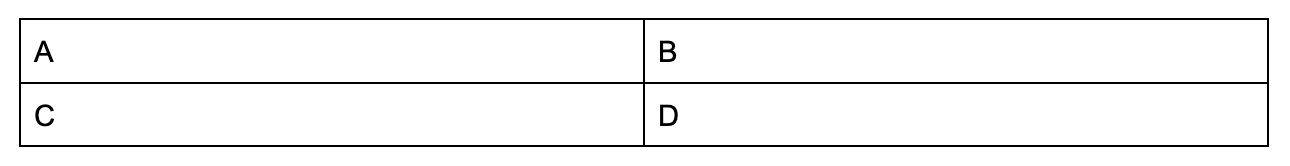}
  \caption{Observed Contingency Table.}
  \label{fig:observed-cont}
\end{figure}

where $A$, $B$, $C$, and $D$ are the entries in the observed contingency table, as shown in Figure \ref{fig:observed-cont}, and $N = A + B + C + D$, which is the total sum of all the elements in the contingency table. This probability is then used to compute a p-value.

When the observed contingency table is of any size other than $2$ x $2$, the Monte-Carlo Method is used. In the Monte-Carlo Method \cite{montecarlo}, we take in the observed contingency table, and generate many other contingency tables with the same row and column sums (we generate 9999, according to the Scipy default \cite{scipymonte}). These tables are generated under the assumption of statistical independence between the 2 groups of qubits being examined. Then, the method evaluates how often the generated tables are at least as extreme as the observed table, under the assumption of independence. Based on this, a p-value is calculated. The p-value is small if the observed table is very unusual, compared to the tables that are generated under the assumption of independence, and the p-value is large if the observed table is consistent with the generated tables.

On the other hand, in the chi-square approach, the chi-square test statistic is calculated through the formula \cite{chi_square_test}:

\[
\chi^2 = \sum_{i=1}^{r} \sum_{j=1}^{c} \frac{(O_{ij} - E_{ij})^2}{E_{ij}}
\]

where $O_{ij}$ and $E_{ij}$ represent the values in the observed contingency table and expected contingency table, respectively. The expected contingency table is derived from the observed contingency table through the formula \cite{chi_square_test}: 

\[
E_{ij} = \frac{R_i \times C_j}{N}
\]

where $R_i$ is the sum of row $i$, $C_j$ is the sum of column $j$, and $N$ is the total sum of all the elements in the table.

In these formulas, we notice that when the observed contingency table has many zeros--which stems from sparse measurement distributions because the observed contingency table is constructed from the counts of the measurement distribution--there will likely be zeros in the expected contingency table, since the row and column sums are likely zero. Then, when computing the chi-square test statistic, from which the p-value is obtained, we notice that when the expected contingency table has zeros, the test statistic will be undefined, since its denominator is the expected contingency table's values. 

Since the measurement distributions of quantum states are commonly sparse, the observed contingency table will commonly have many zeros. Therefore, as explained above, the expected contingency table will also likely have zeros, leading to an undefined chi-square test statistic and thus undefined p-value. However, Fisher's exact test and the Monte Carlo method do not encounter this error due to the approach used in these two statistical tests. Fisher's exact test does not encounter this error because of the formula used to calculate the exact probability. The zeros in the observed contingency table lead the formula's denominator to have a $0!$ term, which is simply $1$, so there will be no division-by-zero error. The Monte Carlo method does not encounter this error because this method does not plug in the observed counts into a specific formula; instead, it compares the observed contingency table to tables that are generated under the assumption of statistical independence \cite{montecarlo}. Therefore, this is why the chi-square test is a worse choice for our purposes than the combination of Fisher's exact test and the Monte Carlo Method.

\subsubsection{Previously Attempted Solutions}
\begin{figure}[t]
  \centering
  \includegraphics[width=\linewidth]{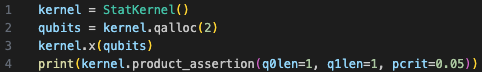}
  \caption{Simple circuit with a single X gate.}
  \label{fig:simpleXcircuit}
\end{figure}
Some prior chi-square-based implementations \cite{yipeng_github} attempt to mitigate this issue by adding 1 to every element in the observed contingency table to eliminate zeros. While this prevents division-by-zero errors, it distorts the observed distribution and produces many small values in the expected contingency table. These small values lead to both degraded performance (extra non-zero entries on which to calculate), and more importantly, poor accuracy because chi-square tests become inaccurate when there are many small expected values in the expected contingency table \cite{chi_square_test}.

A specific example where the chi-square-based assertion produces incorrect results is the circuit shown in Figure \ref{fig:simpleXcircuit}. This circuit simply applies an X gate to a 2-qubit register in the classical state $|00\rangle$. Prior work's chi-square-based product state assertion, which uses a chi-square test and adds 1 to every element in the observed contingency table, returns a very small p-value when executed at line 4 after the X gate, implying that the 2 qubits are statistically dependent and are thus entangled. However, we don't expect these qubits to be entangled because we are only applying an X gate to them. On the other hand, a product state assertion that uses our new approach of a combination of Fisher's exact test and the Monte Carlo method returns a p-value of 1, which correctly implies that the qubits are statistically independent and are thus in a product state. 

It is not surprising that the chi-square-based product state assertion fails on this circuit. This is because the measurement distribution of a qubit in a classical state has a peak at the qubit's value, and is zero everywhere else. Therefore, its observed contingency table will have many zeros. Since prior work's chi-square approach adds 1 to every value in the observed contingency table, the resulting contingency table will have many small values, thus leading it to more likely produce inaccurate results, as we observe in our test case. 

Through extensive testing across a range of quantum circuits, using a combination of Fisher's exact test and the Monte Carlo Method by calling \texttt{scipy.stats.fisher\_exact(cont\_table)} consistently yielded results that aligned with theoretical expectations and known entanglement structures, validating its suitability for this statistical assertion.

\subsubsection{Product State Assertion Usage}
Product state assertions help check whether qubits (or groups of qubits) are entangled or not, which is especially important because entanglement is a key part of many powerful quantum algorithms. For example, in the Quantum Teleportation Circuit shown in Figure \ref{fig:Teleport}, the statistical assertion that should be placed at line 21 is a product state assertion, since we expect the qubits to be entangled at this point. So, we expect the product state assertion to return false, meaning the qubits are not in a product state, and thus are likely entangled.

\section{Integrating with CUDA-Q}
The goal of our debugging workflow is to allow statistical assertions to be invoked at arbitrary points within a quantum circuit, enabling inspection of the quantum state mid-circuit during execution. To achieve this, the assertions must be callable from within the circuit itself and must be supplied with the necessary contextual information, such as the relevant qubits or qubit groups to evaluate.

Implementing this capability differs significantly in CUDA-Q compared to prior Qiskit-based debuggers, primarily due to architectural differences between the two frameworks. In particular, CUDA-Q’s kernel-based programming model, which is designed to facilitate efficient integration with GPUs, introduces unique challenges and design considerations. 

\subsection{CUDA-Q Dynamic Kernel Creation}
\begin{figure}[t]
  \centering
  \includegraphics[height=0.68in]{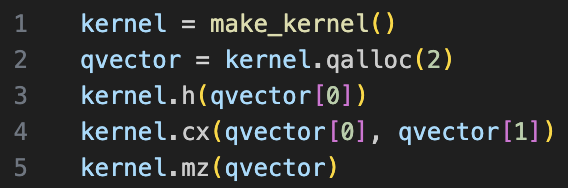}
  \caption{Example of a dynamically-constructed CUDA-Q Kernel.}
  \label{fig:screenshotDynKernel}
\end{figure}

Our approach to integrating the three assertions described in the previous section into CUDA-Q quantum circuits makes use of CUDA-Q's dynamic kernels \cite{nvidia_dynamic_kernel}. These kernels can be instantiated using the \texttt{make\_kernel} function, which creates an empty kernel object of type \texttt{cudaq.PyKernel}. This can be used for dynamic quantum kernel construction outside of a traditional, static kernel function defined as a Python function decorated with \texttt{@cudaq.kernel}. Alternatively, dynamic kernels can also be created directly using the \texttt{PyKernel()} class constructor. This dynamic kernel construction model enables the ability to call quantum operations and construct a quantum circuit outside of a static \texttt{@cudaq.kernel} function. As a result, we can call user-defined classical Python functions in the middle of the kernel. This enables the user to insert our statistical assertions, which are classical Python functions, mid-circuit, thereby allowing them to inspect the circuit at arbitrary points. Figure \ref{fig:screenshotDynKernel} shows an example of a dynamically-constructed kernel.

Specifically, we implemented \texttt{StatKernel}, a custom kernel class that inherits from the \texttt{cudaq.PyKernel} class. This class mirrors the behavior of \texttt{PyKernel}, with one key enhancement: the inclusion of our three statistical assertion methods as part of the class definition. We also provide a modified \texttt{make\_kernel} function, which behaves identically to CUDA-Q’s original one, but returns an instance of \texttt{StatKernel} instead of \texttt{PyKernel}.

To use our debugging workflow, programmers construct their circuits using our version of \texttt{make\_kernel}, which returns a \texttt{StatKernel} instance. This instance supports the invocation of assertion functions at any point within the circuit. For example, we can call \texttt{MyKernel.classical\_assertion} to perform a classical state assertion on the current state of \texttt{MyKernel}. These assertion methods first sample the current quantum state for a specified number of shots using \texttt{cudaq.sample} to construct a measurement distribution, then run the appropriate statistical test on it to check for a specified quantum state (classical, uniform, or product). After the assertion is executed, programmers can continue building the circuit as normal, without explicitly rebuilding the quantum state. Figure \ref{fig:screenshotBV} illustrates statistical assertions in a dynamically constructed circuit.

We choose to use CUDA-Q's dynamic kernels instead of their static kernels, which are kernels defined as Python functions decorated with \texttt{@cudaq.kernel} (e.g. Figure \ref{fig:Teleport}), because user-defined classical Python functions cannot be called from within a static CUDA-Q quantum kernel. Since our statistical assertions are implemented as classical functions, this constraint prevents us from calling them directly inside a kernel. Moreover, CUDA-Q does not support passing classical functions as arguments into kernels, further restricting the ability to inject our assertions mid-circuit. The inability to insert these assertions into CUDA-Q's static kernels to validate the quantum state limits the generality of our debugging tool and represents an avenue for planned future work, especially since CUDA-Q's static kernels are commonly used.

\section{Testing Methodology}
To test our statistical assertions, we evaluated them on a range of quantum circuits. Some circuits were custom-built to specifically target and test the classical, uniform, and product state assertions individually. We also developed test cases that ensure compatibility with quantum kernels that accept arbitrary numbers and types of input parameters.

In addition to these custom tests, we applied our assertion framework to a variety of example algorithms provided in the CUDA-Q documentation \cite{cudaq_all_exs}. These examples include more complex and realistic quantum circuits, allowing us to further validate the robustness and correctness of our implementation in practical scenarios. One challenge encountered during testing with these circuits was having to translate the provided circuits from using static kernels to dynamic kernels, since our assertion tool is designed for dynamic kernels. This translation was not always straightforward due to the fundamental usage differences between the two types of kernels.

\section{Usage Examples}
We now present two examples that demonstrate the application of our debugging tool integrated into real-world quantum programs; they highlight its practical effectiveness in identifying and validating quantum states during circuit execution.

\begin{figure}[t]
  \centering
  \includegraphics[width=\linewidth]{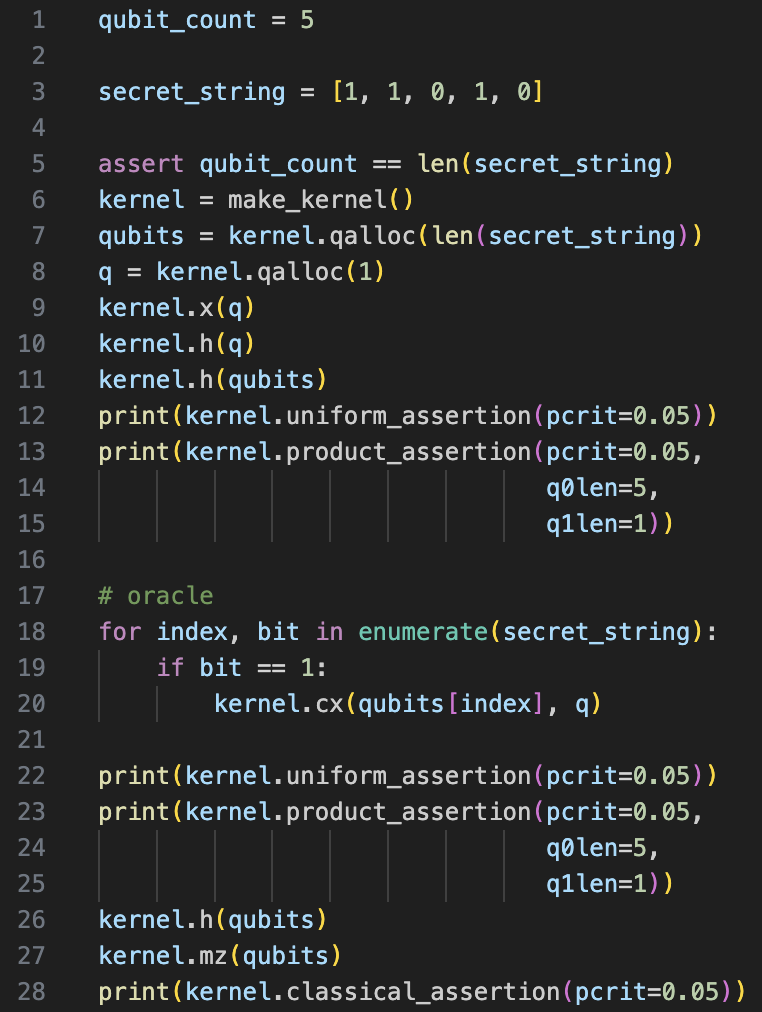}
  \caption{Example of Bernstein-Vazirani Algorithm with assertions.}
  \label{fig:screenshotBV}
\end{figure}

\begin{figure}[t]
  \centering
  \includegraphics[width=\linewidth]{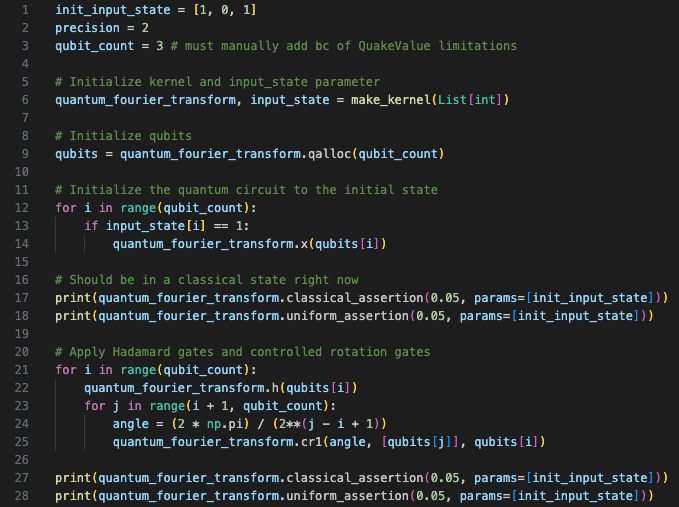}
  \caption{Example of QFT with assertions.}
  \label{fig:screenshotQFT}
\end{figure}

\subsection{Bernstein-Vazirani Algorithm}
Our first usage example is the Bernstein-Vazirani algorithm \cite{bernstein_theory}. This algorithm allows a user to determine a hidden bitstring encoded within an oracle. While a classical approach requires multiple queries to the oracle to determine the bitstring, the quantum algorithm requires only a single query.

\subsubsection{Assertion Placement}
CUDA-Q provides a sample implementation of the Bernstein-Vazirani algorithm \cite{bernstein_cudaq}. However, the provided version is implemented using a static kernel. Therefore, as shown in Figure \ref{fig:screenshotBV}, we translate the original implementation into one that uses a dynamic kernel to enable the use of our debugging assertions. Most of this translation is straightforward, with quantum operations mapping one-to-one. The primary modification involves instantiating the kernel using our \texttt{make\_kernel} function, instead of defining a \texttt{@cudaq.kernel}-decorated function.

After the initial Hadamard gates, but before applying the oracle, we expect the quantum state to be in a uniform superposition that is not entangled. To validate this, we insert both a uniform state and product state assertion at lines 12 and 13 of Figure \ref{fig:screenshotBV}. Notice that the designer specifies the critical p-values for these assertions. However, we also provide support for setting default critical p-value thresholds, since not every QC designer may be familiar with p-values. Then, after applying the oracle, we again expect the state to remain in a uniform superposition that is not entangled, so we place a uniform and product state assertion at lines 22 and 23. Finally, at the very end of the circuit, we expect a classical state, so we insert a classical state assertion here, at line 28. After running CUDA-Q simulations, the outputs of these assertions match with what we expect, validating our debugging workflow.

\subsubsection{Debugging Workflow Example}
To demonstrate the use of our debugging tool, we walk through an example using the Bernstein-Vazirani algorithm code shown in Figure \ref{fig:screenshotBV}. 

One possible bug that could occur in this code is forgetting to apply the Hadamard gate on the \texttt{qubits} register, which contains the secret string, in line 11 of Figure \ref{fig:screenshotBV}. If this occurs, the uniform assertion placed at lines 12 and 22 will return False. However, we expect both of them to be true, according to our theoretical expectations. 

To find the bug, we first identify the earliest point at which the assertion results differ from what we expect, which occurs at the uniform assertion at line 12. This hints that the bug is in the setup code before the oracle. Furthermore, failing a uniform assertion suggests a missing Hadamard gate, since these gates place qubits into a uniform superposition state. Using these hints, we can pinpoint the missing Hadamard gate and fix the bug.

\subsection{Quantum Fourier Transform}
The second usage example we present is the Quantum Fourier Transform (QFT), which is a core quantum subroutine that plays a central role in many notable quantum algorithms, such as Shor’s algorithm. Given the widespread use of QFT in many important algorithms, it is critical to validate its correctness, or else undetected bugs in this subroutine can propagate to a wide range of quantum programs. 

\subsubsection{Assertion Placement}
CUDA-Q provides a sample implementation of the QFT algorithm \cite{qft_cudaq}, but it is expressed using statically-defined kernels. So, we translate this implementation into one that uses dynamic kernels to enable the use of our debugging workflow. We translate like we did with the Bernstein-Vazirani algorithm. However, one new challenge that arises is the need to know the qubit count of the initial state prior to the kernel, due to the lack of native support for passing a \texttt{QuakeValue} variable into a classical function, like \texttt{len}, in a dynamic kernel. Specifically, in the original implementation with a static kernel, \texttt{qubit\_count} can be obtained by calling \texttt{len(input\_state)} within the kernel. However, we cannot do the same with dynamic kernels. Initially, the \texttt{input\_state} parameter that is used within the kernel is passed back to the user as a \texttt{QuakeValue} object after instantiating the dynamic kernel with \texttt{make\_kernel(List[int])}. Then, when we call \texttt{len(input\_state)}, an error is thrown because CUDA-Q cannot handle calling classical functions with \texttt{QuakeValue} objects passed into them as parameters outside of a specified \texttt{@cudaq.kernel}-decorated function, demonstrating a limitation of dynamic kernels in CUDA-Q. However, since we know the initial input state that is passed into the kernel, we can compute its length before kernel construction and store it in the \texttt{qubit\_count} variable, which can then be referenced within the kernel. 

After we translate this QFT implementation from using a static kernel to using a dynamic kernel, we identify critical junctures to insert our assertions. Figure \ref{fig:screenshotQFT} shows the translated QFT circuit with the inserted assertions. We first place a classical state assertion and uniform state assertion immediately after the circuit is initialized with the input state, at lines 17 and 18, since we expect this initial state to be classical. We also place a classical state assertion and uniform state assertion at lines 27 and 28, after we apply the Hadamard gates and controlled rotation gates, since we expect the quantum state here to be in a uniform superposition. These assertions serve as checkpoints to ensure the circuit evolves correctly. 

\subsubsection{Debugging Workflow Example}
One possible bug with this code is forgetting to apply the Hadamard gate at line 22 in Figure \ref{fig:screenshotQFT}. If this bug occurs, our debugging assertions can help to pinpoint this bug because the classical state assertion and uniform state assertion at lines 17 and 18 still return the expected values of True and False, respectively, hinting that the bug is not in the setup code. However, we will notice that the classical state assertion and uniform state assertion at lines 27 and 28 return True and False, respectively, which is not what we expect because the qubits at this point should be in a uniform state. The qubits still being in a classical state here hints that we are probably missing a Hadamard gate after the setup code, therefore helping us to locate the bug.

\section{Future Work}
One important area for future work is optimizing the number of shots used for each assertion. Increasing the number of shots improves the statistical reliability of the assertion results, since we are more confident that our observed measurement distribution is representative of the actual probability distribution, according to the Glivenko–Cantelli Theorem \cite{glivenko_cantelli}. However, shots come at a cost, particularly when running on quantum hardware, since the quantum state must be reconstructed for each shot. Therefore, it is desirable to develop a budget-aware shot allocation strategy that minimizes the number of shots, while still maintaining statistical confidence. 

During the testing of our tool, we noticed that using CUDA-Q's default shot count of 1000 shots would occasionally lead to incorrect results for the product state assertion and uniform state assertion, due to slight fluctuations in their p-value. Increasing the number of shots to 10,000 remediated this issue. On the other hand, decreasing the shot count to 500 did not affect the classical state assertion's accuracy during testing. This aligns with what we expect, since the classical state assertion is testing for a probability distribution with a single large peak, which requires fewer shots to accurately detect than more nuanced, statistically subtle distributions, which is what the uniform and product state assertions assess for. Therefore, budgeting more shots for the uniform and product state assertions and fewer shots for the classical state assertion would be helpful for improving the accuracy of these assertions while minimizing the number of shots used. A more nuanced, mature shot budgeting strategy that can calculate exactly how many shots are needed for a certain level of statistical confidence for each assertion based on the input would be even more helpful because of its adaptability to different inputs, and the customizability it provides to users through being able to customize the statistical confidence levels. 

Another promising direction is extending compatibility to static CUDA-Q kernels. These are more commonly used and have more capabilities, including supporting more modular programming and having fewer issues regarding \texttt{QuakeValue} type compatibility with classical functions. However, integrating statistical assertions into these kernels is difficult due to the fundamental limits of being unable to call user-defined classical Python functions within these kernels. Addressing these challenges would broaden the applicability of our tool and improve its integration into more CUDA-Q workflows.

\section{Conclusion}
In this work, we present a statistical assertion-based debugging workflow for CUDA-Q. In order to promote broader community use, this work is available for open-source use at: \url{https://github.com/jocelynlii/cudaq-assertions}. Our framework improves upon prior QC debuggers by enhancing both usability and the correctness of the product state assertion through the use of a combination of Fisher's exact test and the Monte Carlo Method instead of a chi-square test, while adapting to the unique characteristics of CUDA-Q’s kernel-based programming model. In particular, we note the differences between CUDA-Q's static and dynamic kernels, analyzing how they influence debugger design. Our approach enables more accessible, interpretable, and rigorous debugging for quantum programs developed in CUDA-Q, laying the groundwork for more reliable quantum software development.

\section{Acknowledgments}
This article is based in part upon work supported by the U.S. Department of Energy, Office of Science, National Quantum Information Science Research Centers, and Co-design Center for Quantum Advantage (C2QA) under contract DE-SC0012704.

\bibliography{Bibliography_Jocelyn}

\bibliographystyle{IEEEtran}

\end{document}